\documentclass{article} 
 
\usepackage{amsmath,amssymb}
\usepackage{amsthm}        

\begin{document}

%%%%%%%%%%%%%%%%%%%%% Publisher's Area please ignore %%%%%%%%%%%%%%%
%
%\catchline{}{}{}{}{}
%
%%%%%%%%%%%%%%%%%%%%%%%%%%%%%%%%%%%%%%%%%%%%%%%%%%%%%%%%%%%%%%%%%%%%
\begin{center}
{\bf \Large Finsler geometry as a model for relativistic gravity}
\\[0.2cm]
{\bf Claus L{\"a}mmerzahl
\footnote{ZARM, University of Bremen, 
Am Fallturm, 28159 Bremen, Germany, and  
Institute for Physics, University of Oldenburg, 26111 Oldenburg, Germany
\tt{claus.laemmerzahl@zarm.uni-bremen.de} }
and Volker Perlick
\footnote{ZARM, University of Bremen, 
Am Fallturm, 28159 Bremen, Germany, 
\tt{volker.perlick@zarm.uni-bremen.de} }
}
\end{center}

\begin{abstract}
We give an overview on the status and on the 
perspectives of Finsler gravity, beginning with
a discussion of various motivations for 
considering a Finslerian modification of 
General Relativity. The subjects covered
include Finslerian versions of Maxwell's 
equations, of the Klein-Gordon equation 
and of the Dirac equation, and several
experimental tests of Finsler gravity.

\end{abstract}

\section{Introduction}	

All gravitational phenomena are well described within the theory of General Relativity (GR).
In particular, the Universality of Free Fall, the Local Lorentz Invariance and the Local Position 
Invariance which are at the basis of GR have been experimentally verified with a high 
precision. Together they tell us that the orbit of a pointlike test particle in a  gravitational 
field is uniquely determined by the initial conditions and that all non-gravitational 
experiments give the same result in any local inertial frame irrespective of where and 
when they are carried out \cite{Will93}. This leads to the result that gravity can be 
described by means of a pseudo-Riemannian metric of Lorentzian signature. The question 
of how this metric is related to the energy content of the spacetime is answered by
Einstein's field equation. Einstein arrived at this field equation in 1915 after a long and
arduous process of trial and error. Much later Lovelock \cite{Lovelock1971,Lovelock1972} 
proved that
Einstein's field equation (including a cosmological constant) is uniquely determined 
by the requirements that it contains derivatives of at most second order of the 
metric and that it implies a local conservation law of energy. So if one sticks with metrical
theories of gravity, then there is not much freedom of considering field equations 
other than Einstein's. For tests of Einstein's 
theory in the Solar system and with binary pulsars one usually resorts to the parametrized 
post-Newtonian formalism; until now, all predicted effects have been confirmed 
by observation with high precision \cite{Will:2005va,Lammerzahl:2017wfa}. Einstein's 
theory is also in agreement with observations at the scale of galaxies and clusters of 
galaxies (if one accepts the existence of dark matter) and at cosmological scales (if one 
accepts the existence of dark energy).  

This, however, does not mean that there is no reason for considering possible modifications
of GR. One motivation for such modifications comes from the fact that classical GR and 
Quantum Theory are incompatible. It is usually expected that these two theories should 
merge into a new theory called Quantum Gravity which is still to be found. It is furthermore 
expected that there should be a regime interpolating between GR and Quantum Gravity 
where gravity may be well approximated by a classical (i.e., non-quantum, effective) theory
with small deviations from GR. This interpolating effective gravity theory is generally 
expected to lead to tiny violations of the above-mentioned foundations of GR which may
have two consequences. Firstly, the new theory may have additional gravitational fields 
beyond the pseudo-Riemannian metric as, e.g., scalar fields or a space-time torsion. 
Secondly, the pseudo-Riemannian metric itself may be modified. In particular, it has been
suggested that it may be necessary to replace the pseudo-Riemannian with a Finslerian
metric \cite{Girelleietal2007}. Such a replacement has the consequence that differential
operators such as the d'Alembert operator have to be replaced with pseudo-differential 
operators, thereby introducing non-local features into the theory. This is in agreement
with the general expectation that the implementation of quantum gravity effects leads 
to a non-local theory. 

In this paper we want to review, albeit biased by our personal preferences, 
the present status and the perspectives of Finslerian 
spacetime theories. 
One motivation for our interest in Finsler spacetimes comes from Quantum Gravity,
as indicated above, but this is not the only one. Finslerian geometry can also be used 
for an elegant description of the symmetry given by Very Special Relativity 
\cite{GibbonsGomisPope2007} which, although different from Special Relativity,  
is still compatible with all current experimental limits on violations of Lorentz 
invariance and spatial isotropy. Moreover, a quite different and particularly strong 
motivation for considering Finsler spacetimes comes from the fact that they 
naturally arise from a slight modification of the axiomatic approach to spacetime 
theory by Ehlers, Pirani and Schild \cite{EPS72} as will be outlined in Section \ref{sec:EPS}
below.

%%%%%%%%%%%%%%%%%%%%%%%%%%%%%%%%%%%%%%%%%%%%%%%%%%%%
\section{Mathematical foundation of Finsler geometry}

\subsection{Positive definite case}\label{subsec:podef}
It is worthwile to recall that Riemannian geometry was originally introduced as a 
theory of positive definite metrics, before it was generalised to allow for indefinite 
(`pseudo-Riemannian' or `semi-Riemannian') metrics. Similarly, Finsler geometry 
was originally restricted to the positive definite case. In this version the theory
was brought forward by Paul Finsler in his PhD Thesis of 1919 \cite{Finsler1919},
following a brief remark in Riemann's Habilitation Thesis \cite{Riemann1854}, 
and in this version it is treated in mathematical standard text-books such as the
one by Rund \cite{Rund1959} or by Bao, Chern and Shen \cite{BaoChernShen2000}. We
briefly summarise the main features of positive definite Finsler metrics before turning
to the indefinite case which is of more interest in view of the applications we have in mind.

A positive definite Finsler structure is usually given in terms of a \emph{Finsler
function} $F(x, \dot{x})$ which is defined on the tangent bundle $TM$ of a manifold $M$
with the zero section removed, $(x, \dot{x}) \in TM \setminus \{ 0 \}$. In addition to
being at least three times continuously differentiable, it should be strictly
positive,
\begin{equation}\label{eq:posF}
F(x , \dot{x} ) > 0 \, ,
\end{equation}
positively homogeneous of degree one, 
\begin{equation}\label{eq:homF}
F(x, \lambda  \dot{x}) = \lambda F(x, \dot{x}) \quad 
\text{for \; all \:} \lambda >0 \, ,
\end{equation}
and such that the \emph{Finsler metric}
\begin{equation}\label{eq:metricF}
g_{\mu \nu} (x, \dot{x} ) = 
\dfrac{\partial ^2 F(x,\dot{x})^2}{\partial \dot{x}{}^{\mu} \partial \dot{x}{}^{\nu}}
\end{equation}  
is non-degenerate. Here and in the following we write, by abuse of notation as 
usual, $x=(x^{\mu})$ for a point in the base manifold represented by its coordinates
and $\dot{x}=\big( \dot{x}{}^{\mu} \big)$ for a point in the tangent space at $x$
represented by the induced coordinates. The homogeneity of $F$ implies that 
\begin{equation}\label{eq:gF}
F(x, \dot{x} )^2 = \dfrac{1}{2} \, g_{\mu \nu} \big( x , \dot{x} \big) 
\dot{x}{}^{\mu} \dot{x}{}^{\nu} \, ,
\end{equation}
so the requirement of $F$ being strictly positive implies that the Finsler
metric is positive definite. Moreover, the homogeneity also implies that the 
length functional 
\begin{equation}\label{eq:lengthF}
\ell (x) = \int _a ^b F \big( x(s), \dot{x} (s) \big) ds
\end{equation}
is invariant under reparametrisation for each curve $x : [a,b] \mapsto x(s)$
in $M$. The extremals of the length functional are, by definition, the 
(umparametrised) \emph{geodesics} of the Finsler structure. The Euler-Lagrange
equations of the Lagrangian $L (x , \dot{x} ) = F(x, \dot{x} )^2$ give the
\emph{affinely parametrised geodesics}. 

The Finsler metric is independent of the $\dot{x}$ if and only if
$F(x, \dot{x} )^2$ is a quadratic form. In this case the Finsler structure 
reduces to a Riemannian structure. So roughly speaking Finsler geometry
introduces a way of measuring lengths that is more general than the one
known from Riemannian geometry. In applications of positive definite
Finsler metrics to physics the underlying manifold is to be interpreted
as (three-dimensional) space or as a submanifold thereof. We will now turn
to indefinite Finsler metrics where the underlying manifold may be 
interpreted as (four-dimensional) spacetime.
 
%-----------------------------------------------------------------------------
\subsection{Indefinite case}\label{subsec:indef}
From (\ref{eq:gF}) we read that a generalisation to indefinite Finsler
metrics may be achieved in one of two ways: Either one replaces  
$F( x , \dot{x})^2$ by a function that is allowed to take negative 
values, or one restricts the domain of definition of $F( x , \dot{x})$
to a subset of tangent vectors which are then to be interpreted as 
timelike with respect to the Finsler metric. Both possibilities have been 
worked out in the literature; the first one was pioneered by Beem
\cite{Beem1970}, the second by Asanov \cite{Asanov1985}. 

Beem's definition of a (possibly indefinite) Finsler structure on a 
manifold $M$ is given in terms of a real-valued function $L(x , \dot{x})$
which generalises the square $F(x , \dot{x} )^2$ of the Finsler function.
$L(x, \dot{x})$ is required to be defined and at least three times
continuously differentiable on the tangent bundle with the zero section removed.
It should be positively homogeneous of degree two, 
\begin{equation}\label{eq:homL}
L(x, \lambda  \dot{x}) = \lambda ^2 L(x, \dot{x}) \quad 
\text{for \, all \;} \lambda >0 \, ,
\end{equation}
and such that the \emph{Finsler metric}
\begin{equation}\label{eq:metricL}
g_{\mu \nu} (x, \dot{x} ) = 
\dfrac{\partial ^2 L(x,\dot{x})}{\partial \dot{x}{}^{\mu} \partial \dot{x}{}^{\nu}}
\end{equation}  
is non-degenerate. In analogy to (\ref{eq:gF}), these assumptions imply
that 
\begin{equation}\label{eq:gL}
L(x, \dot{x} ) = \dfrac{1}{2} \, g_{\mu \nu} \big( x , \dot{x} \big) 
\dot{x}{}^{\mu} \dot{x}{}^{\nu} \, .
\end{equation}
As $L$ may take positive or negative values, the Finsler metric may be
indefinite. It must have a specific signature which cannot change from
one point to another because the assumptions guarantee that the determinant 
of the Finsler metric is continuous and nowhere zero. In view of 
applications to physics, we are particularly interested in the case 
that $M$ is four-dimensional and that the signature is Lorentzian, $(+,-,-,-)$.
Mathematically, however, the definition makes sense for any dimension
and any signature.  At each point $x$, we can classify vectors $\dot{x}$ 
as timelike, lightlike or spacelike according to whether $L(x, \dot{x} )$ 
is positive, zero or negative. In the case of Lorentzian signature the
homogeneity assumption guarantees that, at each point $x$, the lightlike 
vectors form a cone which, however, may have more than two connected
components. Criteria for having exactly two connected components (to be
interpreted as a future and a past light cone) have been worked out by
Minguzzi \cite{Minguzzi2015}. 

From a mathematical point of view Beem's definition is quite satisfactory.
In view of physics, however, it is a bit too restrictive because it
excludes several cases which are of interest. Here are two examples.
Firstly, light propagation in a biaxial crystal can be described in terms of 
two Finsler metrics (see Perlick \cite{Perlick2000b}) which, however, violate
Beem's differentiability assumption on a set of measure zero in $TM \setminus \{0\}$.
Secondly, in some static Finsler spacetimes, see Section \ref{subsec:solar} below,
the Lagrangian fails 
to be well-defined on a set of measure zero in $TM \setminus \{0\}$. These 
observations motivated us to relax Beem's definition 
in \cite{LaemmerzahlPerlickHasse2012} by requiring the Lagrangian
to be defined and at least three times continuously differentiable only
\emph{almost everywhere} on $TM \setminus \{0\}$. A stronger 
modification of Beem's definition had been brought forward, already 
earlier, by Pfeifer and Wohlfarth \cite{PfeiferWohlfarth2011}. In addition
to relaxing the regularity conditions in a certain way they also allowed
for positive homogeneity of \emph{any} degree. 

The other approach to Finsler metrics, which is detailed in the book by 
Asanov \cite{Asanov1985}, is restricted to the case of 
Lorentzian signature. Here one sticks with a positive-valued Finsler 
function $F(x , \dot{x} )$ that is positively homogeneous of degree one,
but one restricts its domain of definition, at each point $x$,
to an open conic subset of the tangent space. Asanov calls the vectors
in this domain, which are to be interpreted as timelike, the `admissible
vectors'. In its original version this approach suffered from the 
disadvantage that practically nothing was known about the boundary of 
the domain of admissible vectors, so one did not have any control of 
the vectors one would like to interpret as lightlike. However, this 
disadvantage was overcome in more recent work by Javaloyes and S{\'a}nchez
\cite{JavaloyesSanchez2014} who introduced a refined definition of
Lorentzian Finsler structures in terms of cones and worked out several 
interesting examples.

In view of applications to physics, it seems fair to say that the most 
appropriate definition of indefinite Finsler structures is still a matter 
of debate. We emphasise that here the mathematical details are important.
As a physicist, one usually does not pay much attention to domains
of definition and conditions of differentiability, and in most cases
one gets away with that. In the case at hand, however, such mathematical 
subleties may have a big impact.     

%%%%%%%%%%%%%%%%%%%%%%%%%%%%%%%%%%%%%%
\section{A constructive axiomatic approach to Finsler spacetimes}\label{sec:EPS}

Inspired by earlier work of Reichenbach, Carath{\'e}odory, Weyl and others, 
Ehlers, Pirani and Schild \cite{EPS72} described a constructive axiomatic approach 
to spacetime theory by which the pseudo-Riemannian structure of spacetime and 
gravity is justified and can be tested with a finite number of different experiments. 
This axiomatic approach uses light rays and freely falling particles as the primitive
objects. In the following we briefly review the axioms and we indicate where a 
slight modification leads to a Finslerian spacetime. The observation that a modification
of the Ehlers-Pirani-Schild axiomatics leads to a Finsler structure was made already in
1985 by Tavakol and Van Den Bergh \cite{TavakolVandenbergh1985}. 

The first group of axioms makes spacetime into a differentiable manifold and
the worldlines of light rays and freely falling particles into one-dimensional 
submanifolds. As this part is of no relevance in view of the Finsler modification,
we do not discuss it here. In the next step the conformal structure of spacetime 
is established.  To that end one considers a particle worldline $P$ parametrised with 
a time $t$. This may be just any parametrisation; note that at this stage it does not 
make sense to ask if $t$ is proper time because the latter notion is not yet defined.
An axiom requires that Einstein's synchronisation procedure can be carried through:
For every event $e$ on $P$ there is a neighbourhood $U$ and a 
bigger neighbourhood $V$ such that every event $p$ in $U \setminus P$ can be connected 
to $P$ by exactly two light rays that are contained in $V$. If the events where 
these light rays meet $P$ are denoted $e_1$ and $e_2$, this construction can be 
extended to points $p \in U \cap P$ by setting $e_1=e_2=p$ in this case, thereby defining 
a function 
$g_e : p \mapsto \big( t(e)-t(e_1) \big) \big( t(e_2)-t(e) \big)$ on all of $U$.
The crucial step in this part of the axiomatics is that they required $g_e$ to be 
two times continuously differentiable on all of $U$. In combination with the axioms 
on light propagation this implies that the quantity
\begin{equation}\label{eq:limit}
g_{\mu\nu}(e) = 
\lim_{p \rightarrow e} \frac{\partial^2 g_e(p)}{\partial p^\mu \partial p^\nu}
\end{equation}
is a well-defined symmetric second rank tensor which is 
non-degenerate and of Lorentzian signature. 
Changing the parametrisation on $P$ has the only effect of multiplying $g_{\mu \nu} (e)$ with
a non-zero factor, so this construction defines a conformal equivalence class of Lorentzian 
metrics on the spacetime. From the axioms on light propagation it could be shown that the 
light rays are lightlike geodesics of these Lorentzian metrics.

Since in view of physics differentiability is to be understood as an idealisation that can never
be verified by a finite number of measurements, a postulate of differentiability has to be taken
with care in a constructive axiomatics. To a certain extent, such postulates are necessary and
they had been used in the Ehlers-Pirani-Schild axiomatics 
already in the first step where the differentiable
structure was established. However, such postulates are questionable if relaxing them
leads to a different type of mathematical structure. This is exactly what happens in the case
at hand. If we require $g_e$ to be two times continuously differentiable only on 
$U \setminus P$, the limit on the right-hand side of (\ref{eq:limit}) will in general depend
on the direction from which $p$ approaches the event $e$. As a consequence, one gets
a conformal equivalence class of Finsler metrics. The rest of the Ehlers-Pirani-Schild
axiomatics is about the projective geometry established with the help of 
freely falling particles and about the compatibility of the conformal and the projective 
structure. The details of how to establish a Finsler spacetime precisely in the sense of 
Beem's definition (or some modification thereof) have still to be worked out, but it is
clear that dropping the above-mentioned differentiability postulate in the 
Ehlers-Pirani-Schild axiomatics leads to some kind of Finsler geometry.

%%%%%%%%%%%%%%%%%%%%%%%%%%%%%%%%%%%%%%%%%%%%%%%%%%%%%%%%%%%%%%%%%%%%%%%%%%%%%%%%%%
\section{Finsler gravity}
As motivated in the preceding section, we will now discuss the perspectives of
a Finslerian theory of gravity. We assume that spacetime is a four-dimensional
manifold $M$ and that gravity is coded in an indefinite Finsler structure of 
Lorentzian signature. Unless explicitly referring to another definition, we 
assume that the Finsler structure is defined in terms of a Lagrangian 
function $L(x , \dot{x})$ in the sense of Beem, see Sec.~\ref{subsec:indef}, but 
for the reasons outlined there we require $L(x, \dot{x})$   
to be defined and at least three times differentiable only \emph{almost
everywhere} on $TM \setminus \{0\}$. We use Einstein's summation
convention for greek indices taking value 0,1,2,3 and for latin indices
taking values 1,2,3. Our choice of signature is $(+---)$.
We use units making $\hbar =1$, but we keep the
vacuum speed of light $c$.

If it is possible to find coordinates (on an open neighbourhood $U$ in $M$) 
such that the Lagrangian is independent of $x$, we call
the Finsler structure \emph{flat} (on $U$). In this case we have no
gravitational field but a spacetime that generalises special relativity
in a way that violates Lorentz invariance. 

We will now discuss equations of motion for particles and fields on
a Finslerian spacetime. In the last part of this section we will then
briefly review the various attempts of finding a Finsler generalisation 
of Einstein's field equation. 

%-------------------------------------------------------------------------
\subsection{The geodesic equation}
With the help of the Lagrangian $L(x, \dot{x})$ we define affinely
parametrised geodesics as the solutions to the Euler Lagrange equations
\begin{equation}\label{eq:EL}
\dfrac{d}{ds} 
\Big( \dfrac{\partial L (x, \dot{x} )}{\partial \dot{x}{}^{\mu}} \Big)
- \dfrac{\partial L (x , \dot{x} )}{\partial x^{\mu}} = 0 \, .
\end{equation}
By homogeneity, the Lagrangian $L(x, \dot{x})$ is a constant of motion,
\begin{equation}\label{eq:Lconst}
\dfrac{d}{ds} L (x, \dot{x} ) = 0 
\end{equation}
along every solution of (\ref{eq:EL}). As a consequence we may classify 
geodesics as timelike, lightlike or spacelike according to whether $L(x, \dot{x})$
is positive, zero or negative. Moreover, the fact that $L(x, \dot{x})$ is
homogeneous of degree two implies that a geodesic remains a geodesic under
an affine reparametrisation $s \mapsto as+b$ where $a \neq 0$ and $b$
are real numbers. This is the reason why $s$ is called an `affine parameter'.
Along timelike geodesics we may choose the constant $a$ such that in the 
new parametrisation $L(x , \dot{x}) =1$. In this case the affine parameter
is called (Finsler) \emph{proper time}. 

As an alternative to the Lagrangian formulation, the Finsler geodesics may
also be written in terms of a Hamiltonian. To that end one has to introduce
the canonical momenta
\begin{equation}\label{eq:momenta}
p_{\mu} = \dfrac{\partial L (x , \dot{x} )}{\partial \dot{x}{}^{\mu}}
= g_{\mu \nu} (x, \dot{x} ) \dot{x}{}^{\nu}
\end{equation}
and the Hamiltonian
\begin{equation}\label{eq:Hamilton}
H(x,p) = p_{\mu} \dot{x}{}^{\mu} - L (x, \dot{x} ) =
\dfrac{1}{2} \, g^{\mu \nu} (x, p ) p_{\mu} p_{\nu} 
\end{equation}
where $g^{\mu \nu} (x,p)$ is the contravariant metric, 
\begin{equation}\label{eq:gcontra}
g^{\mu \nu} (x,p) g_{\nu \sigma} (x , \dot{x} ) = 
\delta ^{\mu} _{\sigma} \; .
\end{equation}
In (\ref{eq:Hamilton}) and (\ref{eq:gcontra}) one has to express
$\dot{x}{}^{\mu}$ as a function of $x$ and $p$ with the help
of (\ref{eq:momenta}). 
The timelike, lightlike and spacelike Finsler geodesics are then the solutions 
to Hamilton's equations with $H(x,p)$ positive, zero and negative, respectively. 
As $L ( x , \dot{x})$ is homogenoeus
of degree two with respect to the $\dot{x} {}^{\mu}$, the Hamiltonian
$H(x,p)$ is homogeneous of degree two with respect to the $p_{\mu}$,
hence
\begin{equation}\label{eq:HmuH}
p_{\mu} H^{\mu} (x,p) = 2 \,  H(x,p) 
\end{equation}
where 
\begin{equation}\label{eq:Hmu}
H^{\mu} (x,p) = \dfrac{\partial H(x,p)}{\partial p_{\mu}} \; .
\end{equation}

For the observable features of a Finslerian spacetime structure it is of crucial
importance that timelike geodesics are to be interpreted as freely falling
(massive, structureless) particles and that lightlike geodesics are to be
interpreted as (freely propagating) light rays. This interpretation is a direct
consequence of the axiomatic approach outlined in Sec.~\ref{sec:EPS}. The 
interpretation of lightlike geodesics as light rays can also be justified by
considering the high-frequency limit of appropriately Finsler-modified Maxwell
equations, see next section.  

%------------------------------------------------------------------------------ 
\subsection{Maxwell equations}\label{subsec:maxwell}
In the standard formalism Maxwell's equations, and other field equations,
are partial differential equations for tensor fields on the spacetime
manifold $M$. When generalising to a Finsler setting some features of this
familiar situation have to be given up. Either one has to allow the
fields to live on the tangent bundle $TM$ rather than on $M$, or one 
has to allow the equations to become pseudo-differential equations rather
than differential equations. The first possibility was advertised by Pfeifer and 
Wohlfarth \cite{PfeiferWohlfarth2011}. In their approach, the components
$F_{\mu \nu}$ of the electromagnetic field strength depend not only on
$x$ but also on $\dot{x}$. This requires a rather radical change of the
interpretation of an electromagnetic field which is no longer given in
terms of an invariant geometric object on the spacetime manifold. The
second possibility is more conservative. It was brought forward in the 
appendix of L{\"a}mmerzahl et al. \cite{LaemmerzahlPerlickHasse2012} and
by Itin et al. \cite{ItinLaemmerzahlPerlick2014} and will be briefly
discussed in the following.

We begin by considering a flat Finsler spacetime. We can then find
coordinates such that the Lagrangian and, hence, the Hamiltonian 
(\ref{eq:Hamilton}) is independent of $x$. As a consequence, (\ref{eq:Hmu})
simplifies to
\begin{equation}\label{eq:Hmuflat}
H^{\mu} (p) = \dfrac{\partial H(p)}{\partial p_{\mu}} 
= g^{\mu \nu} (p) p_{\nu}
\; .
\end{equation}
If the space-time metric is the standard Minkowski metric, 
$g^{\mu \nu} = \eta ^{\mu \nu}$ where $\big( \eta ^{\mu \nu} \big) = 
\mathrm{diag} (1,-1,-1,-1)$, Maxwell's equations read
\begin{equation}\label{eq:Max1}
\partial _{\mu} F_{\nu \sigma} + \partial _{\nu} F_{\sigma \mu} 
+ \partial _{\sigma} F_{\mu \nu} \, = \, 0 \, .
\end{equation}
\begin{equation}\label{eq:Max2a}
\eta ^{\rho \sigma} \partial _{\sigma} F_{\rho \nu} = - \mu _0 J_{\nu} 
\end{equation}
where $F_{\rho \nu}$ is the electromagnetic field strength, 
$J_{\nu}$ is the current density and $\mu _0$ is the permeability of 
the vacuum. 

If we replace the Minkowski metric $\eta ^{\rho \sigma}$ with our flat Finsler 
metric $g^{\rho \sigma}(p)$, it is natural to replace
\begin{equation}\label{eq:pdo}
\eta ^{\rho \sigma} \partial _{\sigma} \mapsto 
g^{\rho \sigma} ( - i \partial ) \partial _{\sigma}
\end{equation}
where $i$ is the imaginary unit. Then (\ref{eq:Max2a}) becomes a pseudo-differential
equation,
\begin{equation}\label{eq:Max2g}
g^{\rho \sigma} ( - i \partial ) \partial _{\rho} F_{\sigma \nu} 
= - \mu _0 J_{\nu} 
\end{equation}
whereas (\ref{eq:Max1}) remains unchanged. Eq. (\ref{eq:Max2g}) can also
be written as
\begin{equation}\label{eq:Max2}
i H^{\rho}(-i \partial ) F_{\rho \nu} = - \mu _0 J_{\nu} \, . 
\end{equation}
If the current is given, (\ref{eq:Max1}) and (\ref{eq:Max2}) define a 
perfectly reasonable system of first-order equations for the electromagnetic
field which is a second-rank antisymmetric tensor field on the spacetime
manifold, just as in the standard theory. If the Hamiltonian is specified,
it is an interesting problem to determine the resulting modification of the 
Coulomb potential. This problem was solved in the above-mentioned paper by
Itin et al. \cite{ItinLaemmerzahlPerlick2014} for the case that the metric 
differs from the Minkowski metric by a term of fourth order with respect
to the spatial momentum components. Such a modification of the Coulomb
potential implies, of course, a modification of the hydrogen spectrum,
see below. 

It is not difficult to perform the passage
to ray optics from the equations (\ref{eq:Max1}) and (\ref{eq:Max2}). If one applies
the operator $\partial _{\tau}$ to (\ref{eq:Max2}) for the 
case that $J_{\nu}=0$ and uses (\ref{eq:Max1}), one finds after a bit
of algebra that the electromagnetic field strength satisfies a 
generalised wave equation,
\begin{equation}\label{eq:wave}
H( - i \partial ) F_{\mu \nu} = 0 \, .
\end{equation}
This equation is solved by a plane-wave ansatz 
\begin{equation}\label{eq:pw}
F_{\mu \nu}(x) = 
\mathrm{Re} \Big\{f_{\mu \nu} \, \mathrm{exp}(ik_{\sigma}x^{\sigma}) \Big\} \, ,
\end{equation}
only if the wave covector $k_{\sigma}$ satisfies the equation
\begin{equation}\label{eq:eikonal}
H( k )  = 0 \, ,
\end{equation}
which demonstrates that on our flat Finsler spacetime 
electromagnetic waves propagate along lightlike straight lines.

On a curved Finsler spacetime the partial derivatives in (\ref{eq:Max2})
and, thus, in (\ref{eq:wave}) have to be replaced by some kind of
covariant derivatives to give a coordinate independent meaning to
these equations. By a generalisation of the above argument one can 
then show that on a curved Finsler spacetime high-frequency 
electromagnetic waves propagate along lightlike geodesics, see the 
Appendix of L{\"a}mmerzahl et al. \cite{LaemmerzahlPerlickHasse2012}. 

%--------------------------------------------------------------------------
\subsection{Klein-Gordon equation}
In analogy to Maxwell's equations, the Klein-Gordon equation can
be generalised into a Finsler setting in two quite different ways: In the
first approach, which was advertised in particular by Asanov 
\cite{Asanov1985}, one allows the field to depend not only on the 
$x^{\mu}$ but also on the $\dot{x}{}^{\mu}$; the Klein-Gordon 
equation is then a differential equation involving a generalised wave 
operator which also differentiates with respect to the $\dot{x}{}^{\mu}$. 
In the second approach one leaves the field, as in the standard theory, to 
depend on the $x^{\mu}$ only and allows the Klein-Gordon equation
to become a pseudo-differential equation. We will here follow the second
approach.

As in the case of Maxwell's equations, we first consider a flat 
Finsler spacetime given, in appropriately chosen coordinates, by 
a Hamiltonian that is independent of $x$,
\begin{equation}\label{eq:Hflat}
H(p)= \dfrac{1}{2} g^{\mu \nu} (p) p_{\mu} p_{\nu} \, .
\end{equation}
If the metric is the usual Minkowski metric, $g^{\mu \nu} =
\eta ^{\mu \nu}$, the Klein-Gordon equation for a complex-valued
scalar field $\Phi$ with mass parameter $m$ reads
\begin{equation}\label{eq:KGMink}
\eta ^{\rho \sigma} \partial _{\rho} \partial _{\sigma} \Phi
+ m^2 \Phi = 0 \, .
\end{equation}
If we replace the Minkowski metric $\eta ^{\rho \sigma}$ with our flat 
Finsler metric $g^{\rho \sigma}(p)$, we replace the wave operator in 
(\ref{eq:KGMink}) according to the same rule as in (\ref{eq:pdo}), 
\begin{equation}\label{eq:pdo2}
\eta ^{\rho \sigma} \partial _{\rho} \partial _{\sigma} \mapsto 
g^{\rho \sigma} ( - i \partial ) \partial _{\rho} \partial _{\sigma} \, .
\end{equation}
This gives us the Finslerian version of the Klein-Gordon equation,
\begin{equation}\label{eq:KGFin}
g^{\rho \sigma} ( - i \partial ) \partial _{\rho} \partial_{\sigma} 
\Phi + m^2 \Phi = 0  
\end{equation}
which can also be written as
\begin{equation}\label{eq:KGFin2}
2 \, H ( - i \partial ) \Phi + \, m^2 \Phi = 0 \, .  
\end{equation}
Clearly, (\ref{eq:KGFin2}) is a pseudo-differential equation for the 
scalar field $\Phi$. 

The non-relativistic limit of (\ref{eq:KGFin2}) gives a Finsler-modified
free Schr\"odinger equation. This has been worked out by Itin et al.
\cite{ItinLaemmerzahlPerlick2014} for the case of a flat Finsler
metric that differs from the Minkowski metric by terms of fourth order
in the spatial momentum coordinates. In the same paper, the 
free Schr{\"o}dinger equation was then replaced by the 
Schr{\"o}dinger equation with a Finsler-modified
Coulomb potential which allowed calculating Finsler perturbations of 
the hydrogen atom. We will come back to this work in Section
\ref{subsec:LI} below.  

On a curved Finsler spacetime, (\ref{eq:KGFin}) has to be modified in 
two ways. Firstly, the Hamiltonian is then a function not only on
the $p_{\mu}$ but necessarily also of the $x^{\mu}$. Secondly, one
needs to make the differential operator coordinate-independent by
adding terms involving Finslerian Christoffel symbols. Therefore,
the resulting equation is of the form
\begin{equation}\label{eq:KGFin3}
2 \, H ( x , - i \partial ) \Phi + \, m^2 \Phi + \, \ldots \, = 0 
\end{equation}
where the ellipses indicate a term involving a pseudo-differential
operator that is homogeneous of degree one acting on $\Phi$.

%-----------------------------------------------------------------------
\subsection{Dirac equation}

A Finsler generalisation of the Dirac equation is not straightforward 
since there are conceptually different approaches. This is related to 
how the transition from the Klein-Gordon equation to the Dirac equation
is performed. In the following we discuss in some detail two different
routes from the Finslerian Klein-Gordon equation to a Finslerian 
Dirac equation. We restrict to the case of a flat Finsler spacetime
before commenting, at the end of this section, on the case of a 
curved spacetime.   

\subsubsection{Reducing the Finslerian Klein-Gordon equation to first order}

We first take the route which starts from the Klein-Gordon equation
(\ref{eq:KGFin2}) on a flat Finsler spacetime and ask for a related 
first-order differential equation of the form
\begin{equation}\label{eq:Dirac1}
0 = \gamma(- i \partial) \psi + m \psi \, .
\end{equation}
Here we consider a field variable $\psi(x) \in \mathbb{C}^r$, for some 
$r \in \mathbb{N}$, and $\gamma (p)$ is an $r \times r$ matrix that 
is positively homogeneous  of degree one, $\gamma(\lambda p) = 
\lambda \gamma(p)$ for $\lambda > 0$. In the following we write
\begin{equation}\label{eq:gammamu}
\gamma ^{\mu} (p) = \dfrac{\partial \gamma (p)}{\partial p_{\mu}}
\, , \quad
\gamma ^{\mu \nu} (p) = 
\dfrac{\partial ^2 \gamma (p)}{\partial p_{\mu} \partial p_{\nu}}
\, .
\end{equation}
Owing to the homogeneity (\ref{eq:Dirac1}) can be rewritten as 
\begin{equation}
0 = - i \gamma^\mu(- i \partial) \partial_\mu \psi + m \psi \, , 
\end{equation}
which looks like the ordinary Dirac equation but with generalised
Dirac matrices $\gamma ^{\mu} (p)$ that are homogeneous of 
degree zero with respect to the $p_{\mu}$. The compatibility with 
(\ref{eq:KGFin2}) then requires 
\begin{equation}\label{eq:Hgamma}
\gamma^2(p) = H(p) \, \boldsymbol{1},
\end{equation}
where $\boldsymbol{1}$ is the $r \times r$ unit matrix.
This equation can be rewritten as $\gamma^\mu(p) \gamma^\nu(p) p_\mu p_\nu 
= 2 g^{\mu\nu}(p) p_\mu p_\nu \boldsymbol{1}$. 
Because of the $p$-dependence of the 
$\gamma ^{\mu}$ it is clear that the usual Clifford algebra has to be 
modified, 
\begin{equation}
\gamma^\mu(p) \gamma^\nu(p) + \gamma^\nu(p) \gamma^\mu(p) 
+ \gamma^{\mu\nu}(p) \gamma(p) + \gamma(p) \gamma^{\mu\nu}(p) 
= 2 g^{\mu\nu}(p) \, \boldsymbol{1} \, .
\end{equation} 
Only if $\gamma$ depends linearly on $p$ do we recover the usual Clifford algebra. 

For a given Finsler Hamiltonian $H(p)$ it is possible to calculate the $\gamma$
matrices in terms of a series expansion with respect to the Finslerian deviation 
from the Minkowski space Hamiltonian $H_0(p) = \frac{1}{2} \eta ^{\mu \nu} 
p_{\mu} p_{\nu}$. We work this out for the special case that $H(p)$ is of the 
form
\begin{equation}\label{FinslerPolynom1}
2 H(p) = p_0^2 - \Big( \big( 
\delta ^{i_1i_2} \cdots \delta ^{i_{2n-1}i_{2n}} + \psi ^{i_1 \ldots i_{2n}} \big)
p_{i_1} \cdots p_{i_{2n}} \Big) ^{1/n} 
\end{equation}
with some integer $n$. The $\psi ^{i_1 \ldots i_{2n}}$ are the components of
a purely spatial symmetric tensor of rank $2n$. Upon writing the Hamiltonian as
\begin{equation}\label{FinslerPolynom2}
2 H(p) = p_0^2 - \big| \vec{p} \big| ^2 \left( 
1 + \dfrac{\psi \big( \vec{p} \big)}{\big| \vec{p} \big|^{2n}} \right) ^{1/n}
\end{equation}
where
\begin{equation}\label{eq:psi}
\big| \vec{p} \big| ^2 = \delta ^{ij} p_i p_j \, , \quad
\psi \big( \vec{p} \big) = \psi ^{i_1 \ldots i_{2n}} p_{i_1} \cdots p_{i_{2n}} \, ,
\end{equation}
Taylor expansion yields
\begin{equation}\label{eq:HTaylor}
2 H(p) \! = \! p_0^2 \!  - \! \big| \vec{p} \big| ^2 \! \left( 
1 + \dfrac{1}{n} \dfrac{\psi \big( \vec{p} \big)}{\big| \vec{p} \big|^{2n}}
+ \dfrac{(1-n)}{2! n^2} 
\Big( \dfrac{\psi \big( \vec{p} \big)}{\big| \vec{p} \big|^{2n}} \Big)^2
+ \dfrac{(1-n)(1-2n)}{3! n^3}  
\Big( \dfrac{\psi \big( \vec{p} \big)}{\big| \vec{p} \big|^{2n}} \Big)^3
\ldots \right)
\end{equation}
We want to solve (\ref{eq:Hgamma}) with the ansatz
\begin{equation}\label{eq:ansgamma}
\gamma(p) = \gamma^\mu(p) p_\mu = \gamma^\mu_{(0)} p_\mu + \gamma^\mu_{(1)}(p) p_\mu + \gamma^\mu_{(2)}(p) p_\mu + \ldots
\end{equation} 
where the $\gamma^\mu_{(s)}(p)$ are of the $s$th order in $\psi$. Inserting (\ref{eq:ansgamma}) 
and (\ref{eq:HTaylor}) into (\ref{eq:Hgamma}) and comparing terms of equal order in $\psi$ gives
a hierarchy of equations that can be solved successively. A first step in this direction was taken 
in \cite{Zimmermann17}. 

Though being quite natural and straightforward this is a rather cumbersome approach. In principle,
it works for every Finsler Hamiltonian that admits a Taylor expansion about the Minkowski Hamiltonian.
The resulting Dirac equation is, in general, a pseudo-differential equation for an $r$-component
(spinor) field whose components are functions only of the $x^{\mu}$. We will now briefly 
outline another approach which gives us a differential equation rather than a pseudo-differential
equation, but it works only for very special Finslerian structures..

\subsubsection{Reducing a higher-order scalar differential equation to first order}

For the second route we start out from a Klein-Gordon-like equation 
of the form  
\begin{equation}
\eta^{\mu_1 \ldots \mu_{2n}} \partial_{\mu_1} \cdots \partial_{\mu_{2n}} \Phi 
= \left(\frac{m^2}{2}\right)^n \Phi 
\end{equation}
where $n$ is a positive integer and $\eta^{\mu_1 \ldots \mu_{2n}}$ is a symmetric
tensor of rank $2n$.  
We may think of this equation as coming from a Finsler structure in the sense 
of Pfeifer and Wohlfarth \cite{PfeiferWohlfarth2011} who allow for homogeneity
of any degree. We ask the question of whether this higher-order equation can 
be reduced to a first-order equation which has the form of a generalised Dirac 
equation, $i \gamma^\mu \partial_\mu \psi - m \psi = 0$. Here 
we mean by `reduced' that any solution of the first-order generalised Dirac 
equation is also a solution of the higher-order equation. A $2n$ fold iteration 
of the Dirac equation then leads to the condition
\begin{equation}
\gamma^{(\mu_1} \gamma^{\mu_2} \cdots \gamma^{\mu_{2n})} 
= \eta^{\mu_1 \ldots \mu_{2n}} \boldsymbol{1} \label{GCA}
\end{equation}
which clearly is a generalisation of the standard Clifford algebra for $n = 1$. 

In the positive definite case, such systems have been discussed in the literature. 
Roby \cite{Roby1969} used the concept of a generalised Clifford algebra for the 
linearisation of $n$-forms. Nono \cite{Nono71} discussed the linearisation of 
higher-order equations $g^{\mu_1 \ldots \mu_m} \partial_{\mu_1} \cdots 
\partial_{\mu_m} \Phi = c^m \Phi$ which is similar to our approach. He also 
derived the condition (\ref{GCA}) and also a slightly more general condition. 
A short review on various concepts of generalised Clifford algebras has been 
given by Childs \cite{Childs1978}. 

\subsubsection{Other approaches and generalisations to curved Finsler spacetimes}
In addition to the two routes sketched above, other approaches to a Finslerian Dirac equation
have been suggested. For the case of flat Finsler spacetimes, to which we have restricted
here, we mention that Finsler-type modifications of Special Relativity naturally arise in
the Standard Model Extension, see e. g. Kosteleck{\'y} \cite{Kostelecky2011}. In this
context modified Dirac equations have been discussed by various authors, see e.g.
Lehnert \cite{Lehnert2004}. Moreover, several authors have considered modified Dirac
equations on curved Finsler manifolds. To the best of our knowledge, the first
attempt of formulating a Finsler version of the Dirac equation was brough forward by
Asanov \cite{Asanov1985}. His Dirac spinors depend on two variables where the first one
ranges over the spacetime manifold and the second one is related to a parametric 
representation of the indicatrix $L=1$. Asanov calls this the `parametric representation
of physical fields'. Bogoslovsky and Goenner \cite{BogoslovskyGoenner2004} discussed a 
Finslerian metric which is conformally related to the Minkowski metric. Within this framework 
they also proposed a Dirac equation. The $\gamma$-matrices are still the usual ones; the 
modifications come in through the modified transformations of the vectors and spinors.
Finally, we mention the work by Vacaru, see e.g. \cite{Vacaru2006}, on Clifford-Finsler 
algebroids and modified Dirac equations.

%----------------------------------------------------------------------
\subsection{Generalised Einstein equation}
Until now we have assumed that a Finslerian spacetime is 
given and we have discussed equations of motion for particles
or fields on this spacetime. Now we have to address the 
question of how the Finslerian spacetime (i.e.,
the gravitational field) is determined by the distribution of 
energy, i.e., how Einstein's field equation has to be modified
to fit into a Finslerian setting.  Einstein's field equation is of 
the general form that the curvature of spacetime is algebraically 
related to its energy content. If one wants to preserve this
general form one faces a major problem: In Finsler geometry
there are various different curvature tensors but none
of them lives on the base manifold; if written in local coordinates, 
the components of the curvature tensor are functions not only 
of the $x^{\mu}$ but also of the $\dot{x}{}^{\mu}$. Therefore,
a straight-forward way of generalising Einstein's field equation
would require also the energy content of the spacetime to be
described by an object that depends on the $x^{\mu}$ and on
the $\dot{x}{}^{\mu}$, i.e., by an energy-momentum tensor
that lives on the cotangent bundle. This would mean a major
change in interpretation: We are used to modelling the 
energy content of a spacetime in terms of tensor or spinor fields
on the spacetime; letting the fields depend on the 
$\dot{x}{}^{\mu}$ would be a way of saying that they 
are \emph{not} such invariant (i.e., coordinate independent
and observer independent) fields on the spacetime. We emphasise
that this problem occurs only at the level of the gravitational
field equation: As long as we consider (Maxwell, Klein-Gordon, Dirac
... ) fields on a Finsler background spacetime, without taking the
self-gravity of these fields into account, we may very well
consider these fields as tensor or spinor fields on the spacetime,
as demonstrated in the preceding sections.

Several different Finsler generalisations of Einstein's field
equation have been suggested, but it seems fair to say that
up to now none of them is generally accepted. To the best
of our knowledge, the first such generalisation was brought
forward by Rund and Beare \cite{RundBeare1972} in 1972. 
Their set-up is based on curvature and energy quantities that
depend on the $x^{\mu}$ and on the $\dot{x}{}^{\mu}$,
as indicated above. Since the authors were not able to formulate
a law of local energy conservation, as it is valid in the
standard formalism, they themselves doubted that their
approach gives a physically reasonable generalisation  of 
Einstein's field equation. 

A few years later, Asanov 
\cite{Asanov1983} made a different suggestion. He reduced
the curvature quantities from the tangent bundle over
spacetime to the spacetime itself by what he called the
notion of `osculation': He chose an auxiliary timelike
vector field $V^{\mu} (x)$ and replaced, in the argument 
of the curvature quantities, the $\dot{x}{}^{\mu}$ with
$V^{\mu} (x)$. The problem with this approach is, of
course, that the geometry of spacetime is not invariant 
but depends on a vector field that may be interpreted, at 
each point, as the four-velocity of an observer. Such an 
observer-dependent geometry is very much against the 
spirit of general relativity. 

A completely different approach was suggested by Rutz
\cite{Rutz1993}. She restricted herself to the question
of how the \emph{vacuum} Einstein equation could be
generalised into a Finslerian setting. As a guiding principle,
she used the idea that the vacuum field equation should 
express the fact that the tidal tensor be trace-free. This
is true in Newtonian theory, where the tidal tensor is 
the Hessian of the Newtonian potential, and also in 
Einstein's theory, where the tidal tensor is the Riemannian
curvature tensor. Consequences of the resulting vacuum
equation have been discussed in some detail. In particular
a plethora of spherically symetric solutions has been found, 
see Rutz \cite{Rutz1998}. The generalisation to the matter 
case, however, remains an open problem.  

Another version of a vacuum field equation, in this
case for Finsler spacetimes with a certain product structure,
was inverstigated in several papers by Vacaru, see for 
instance \cite{Vacaru2013}. In particular, black-hole
solutions of this field equation have been worked out.    

Finally, we mention the work by Pfeifer and 
Wohlfarth \cite{PfeiferWohlfarth2012} who brought 
forward a Finsler generalisation of Einstein's field equation,
based on their definition of Finsler spacetimes which is a 
generalisation of Beem's, see Section \ref{subsec:indef}. These
authors decidedly take the view that curvature and energy quantities 
should, indeed, depend on the $x^{\mu}$ and on the 
$\dot{x}{}^{\mu}$, and they discuss the corresponding 
notions of observers and measurements in some detail. Although
their approach is certainly satisfactory from a mathematical
point of view, we believe that there are still open questions 
in view of the physical interpretation. Therefore, in our view,
the problem of finding a Finsler generalisation of Einstein's
field equation is still open.
 
%%%%%%%%%%%%%%%%%%%%%%%%%%%%%%%%%%%%%%%%%%%%%%%%%%%%%%%%%%%%%%%%%%%%%%%%%%%%
\section{Experimental tests}

\subsection{Finslerian violation of Lorentz invariance}\label{subsec:LI}
In sufficiently small spacetime regions we may neglect gravity, 
i.e.,we may approximate the spacetime metric by a flat metric.  
At this level of approximation, a hypothetical Finsler modification
of spacetime theory comes up to replacing the Minkowski metric 
of special relativity with a flat Finsler metric. A characteristic feature
of such a modification is a violation of Lorentz invariance, in particular
of spatial isotropy, which can be experimentally tested in various ways.

One example is the test of anisotropies in the propagation of light 
with the help of Michelson interferometry, see  
L{\"a}mmerzahl et al. \cite{LaemmerzahlLorekDittus2008}.
If optical resonators are used instead of traditional Michelson 
interferometers, the isotropy of the velocity of light has been 
verified with extremely high accuracy. This gives very strong 
limitations on Finsler perturbations based, however, on the 
asssumption that only the light propagation but not 
the length of the resonator (or of the interferometer 
arms) is affected by the Finsler perturbation in a measurable 
way. In the above-mentioned paper arguments are given why this 
assumption is, indeed, justified. 

As another possibility, Finslerian anisotropies may also be observed
with the help of spectroscopy. We have seen in Section 
\ref{subsec:maxwell} that in a Finsler spacetime the Coulomb law 
will be modified. As a consequence, the energy levels of the
hydrogen atom will change. We have worked this out for a 
flat Finsler metric given by a Hamiltonian $H$ of the form
\begin{equation}\label{eq:Hflat1}
2 \, H (p)=  
p_0^2 - 
\sqrt{\big( \delta^{ij} \delta ^{kl} + \psi ^{ijkl} \big)
p_i p_j p_k p_l } \, ,
\end{equation}
see Itin et al. \cite{ItinLaemmerzahlPerlick2014}. Here
$\psi ^{ijkl}$ is a totally symmetric spatial fourth-rank 
tensor that describes a Finsler perturbation of
the Minkowski metric. Assuming that the $\psi ^{ijkl}$
are so small that all equations can be linearised with
respect to them, the Hamiltonian can be simplified 
by a linear coordinate transformation to the form
\begin{equation}\label{eq:Hflat2}
2 \, H (p)=  
\eta ^{\mu \nu} p_{\mu} p_{\nu} - 
\dfrac{2  \phi ^{ijkl} p_i p_j p_k p_l }{\delta ^{mn}p_mp_n} 
\end{equation}
where the redefined Finsler perturbation tensor $\phi ^{ijkl}$ is 
totally symetric and trace-free, so there are 9 independent 
components. After determining the Finsler modified Coulomb 
potential we have set up the Finsler modified Schr{\"o}dinger 
equation and calculated the eigenvalues for the quantum numbers
$n=1$, $n=2$ and $n=3$ of the Hamiltonian
with the help of perturbation theory.  
In general, the Finsler coefficients give rise to a splitting of
the Lyman-$\alpha$ line (transition from $n=2$ to $n=1$) 
and of the Lyman-$\beta$ line (transition from $n=3$ to $n=1$).
If we observe, with a measuring accuracy $\delta \omega$ of the 
frequency, that these two lines do not split, our calculated 
values of the shifts lead to upper bounds on the $|\phi ^{ijkl}|$
in the order of $10^{-17} \delta \omega / \mathrm{Hz}$.
As frequencies can be measured in the optical and in the 
ultraviolet with an accuracy of up to $\delta \omega \approx 10^{-7}
\mathrm{Hz}$, we see that atom spectroscopy gives us bounds on 
the Finsler coefficients in the order of $10^{-24}$. Nuclear 
spectroscopy might give even smaller bounds, but this has not 
been worked out until now.

%----------------------------------------------------------------------------------------------
\subsection{Solar system tests of Finsler gravity}\label{subsec:solar}
The PPN formalism, which is routinely used as a mathematical framework
for modelling possible deviations from Einstein's theory, is restricted to
theories where the gravitational field is described in terms of a 
pseudo-Riemannian metric. However, similar post-Newtonian expansions
have also been developped for special Finsler metrics which have been
suggested as hypothetical Finslerian models of the Solar system. 
In particular, 
Roxburgh \cite{Roxburgh1992a} used such expansions for a certain 
quartic Finsler metric. In another paper \cite{Roxburgh1992b} he
considered a Finsler metric that differs from a 
pseudo-Riemannian metric only by a (nowhere vanishing) scalar factor;
such Finsler spacetimes are, of course, very special because
they have the same lightlike geodesics as a pseudo-Riemannian
metric, so the laws of light deflection are unaffected by this kind of
Finsler modification. 

For this reason, we suggested a different mathematical setting for 
Solar system tests of Finsler gravity, see L{\"a}mmerzahl et al.
\cite{LaemmerzahlPerlickHasse2012}. In spirit it is similar to 
the PPN formalism but the mathematical technicalities are different.
We start out from a Finsler spacetime with a Lagrangian $L$ of the 
form
\begin{equation}\label{eq:defL}
2 \, L=  
\big( h_{tt} + c^2 \psi _0 \big) 
{\dot{t}}{}^2 - \Big( \big( h_{ij} h_{kl} +
\psi _{ijkl} \big) {\dot{x}}{}^i {\dot{x}}{}^j {\dot{x}}{}^k {\dot{x}}{}^l
\Big)^{\frac{1}{2}} \; .
\end{equation}
Here 
\begin{equation}\label{eq:defh}
h_{tt} dt^2 - h_{ij} dx^i dx^j =  
\big( 1 - \frac{\, 2 \, G \, M \,}{c^2 \, r} \big) c^2 dt^2 -  
\dfrac{dr^2}{1 - \frac{\, 2 \, G \, M \,}{c^2 \, r} }
-  r^2 \big( {\mathrm{sin}}^2 \vartheta \,
d \varphi ^2 + d \vartheta ^2 \big)
\end{equation}
is the Schwarzschild metric, the spatial perturbation tensor 
field $\psi _{ijkl}$ is spherically symmetric and independent 
of $t$, and the time perturbation
$\psi _0$ is a function of $r$ only. The fourth-order term $\psi _{ijkl}  
{\dot{x}}{}^i {\dot{x}}{}^j {\dot{x}}{}^k 
{\dot{x}}{}^l$ may be viewed as the leading order term in 
a general Finsler power--law perturbation of the spatial  part 
of the metric. Note that (\ref{eq:defL}) is an
example of a Finsler Lagrangian which does not satisfy Beem's 
definition because it is not defined and  three times
continuously differentiable on all of 
$TM \setminus \{ 0 \}$: It is not well behaved on vectors tangent
to a $t$-line. The fact that we want to include static metrics of
this form is the main motivation why we relaxed Beem's definition
by requiring $L$ to be defined and three times continuously 
differentiable only \emph{almost everywhere} on $TM 
\setminus \{ 0 \}$, recall Section \ref{subsec:indef}.

We assume that the Finsler perturbation is so small that we may
linearise all expressions with respect to $\psi _{ijkl}$ and
$\psi _0$. Moreover, because of the spherical symmetry we
may restrict to the equatorial plane $\vartheta = \pi /2$ when
discussing timelike and lightlike geodesics. After an appropriate
transformation of the radius coordinate the Lagrangian can then 
be rewritten as
\begin{equation}\label{eq:linL}
2 L = (1+ \phi _0)  
h_{tt} {\dot{t}}{}^2 - 
(1+\phi_1) h_{rr} {\dot{r}}{}^2
- r^2 {\dot{\varphi}}{}^2 -
\dfrac{\phi _2 h_{rr} r^2 {\dot{r}}{}^2  {\dot{\varphi}}{}^2
}{
h_{rr}  {\dot{r}}{}^2 + r^2 {\dot{\varphi}}{}^2 } 
\end{equation}
with redefined Finsler perturbations $\phi _0$, $\phi _1$ and 
$\phi_2$ which are functions of $r$ only. Note that $\phi_0$
just changes the time measurement while $\phi _1$ 
changes the radial length measurement, i.e., these two
perturbations just lead to a modified pseudo-Riemannian
metric. By contrast, $\phi _2$ describes a genuine Finsler
perturbation. We call $\phi _2$ the `Finslerity'. 

In L{\"a}mmerzahl et al. \cite{LaemmerzahlPerlickHasse2012}
we calculated the orbits of timelike and lightlike geodesics
in the geometry given by (\ref{eq:linL}). Considering this
geometry as a model for the Solar system, this allowed us 
to determine the effect of the Finsler perturbation on the 
perihelion precession of planets and on the time delay and 
the deflection of light. It was our main goal to find 
observational bounds on the Finslerity.  Assuming that
the Finsler perturbations have a fall-off behaviour as
\begin{equation}\label{eq:falloff}
\phi _A (r) \, = \, \phi _{A1} \, \dfrac{2GM}{c^2r} \, + \,
O \Big( \big( \dfrac{2GM}{c^2r} \big) ^2 \Big) 
\, , \qquad A=0,1,2 \, ,
\end{equation}
we found from Solar system observations that
\begin{equation}\label{eq:boundphi21}
\big| \,  \phi _{21} \, \big| \, \lessapprox \, 10^{-3} \; .
\end{equation}
This bound is surprisingly weak, much weaker than the
bounds on Finsler perturbations from atom
spectroscopy, cf. Section \ref{subsec:LI}.

We conclude this section with the remark that tests of 
GR with Solar system ephemerides are usually based
on the PPN formalism and do not cover Finsler
perturbations. It might be worthwile to include
some kind of `Finsler parameter' into these
considerations.

%----------------------------------------------------------------------------------------------
\subsection{Redshift experiments}\label{subsec:redshift}
If light is emitted with a certain frequency $\omega _1$
by an observer, it will in general arrive with a different 
frequency $\omega _2$ when received by another observer.
The quantity 
\begin{equation}\label{eq:z}
z = \dfrac{\omega  _1 - \omega _2}{\omega _2}
\end{equation}
is called the redshift. Here it is understood that $\omega _1$
and $\omega _2$ are measured with respect to standard
clocks by the respective observer. In general, $z$ comes
about as a combination of effects from the relative motion
(Doppler shift) and from the spacetime geometry (gravitational
redshift). Redshift experiments are appropriate for testing
spacetime theories on Earth, in the Solar system and at 
cosmological scales.

Obviously, redshift experiments crucially depend on the
notion of standard clocks. On a Finsler spacetime, a standard
clock is defined as a clock that parametrises its worldline
with a parameter $\tau$, called (Finsler) \emph{proper time}, 
according to
\begin{equation}\label{eq:proper}
g _{\mu \nu} ( \gamma ( \tau ) , \dot{\gamma} ( \tau ) ) 
\dot{\gamma}{}^{\mu} \, \dot{\gamma}{}^{\nu} = 1 \, .
\end{equation}
Here $g_{\mu \nu}$ denotes the Finsler metric and $\gamma ^{\mu}
( \tau )$ is the parametrised worldline. Einstein's synchronisation
procedure can then be used in the usual way for assigning 
a \emph{radar time} and a \emph{radar distance} to events in 
a neighbourhood of a standard clock, cf. Pfeifer \cite{Pfeifer2014}.

For any two worldlines parametrised with proper time, $\gamma _1
( \tau _1)$ and $\gamma _2 ( \tau _2 )$, the redshift can be 
written as
\begin{equation}\label{eq:redsh3}
1+z= 
\dfrac{
g_{\mu \nu} \big( x ( s_1) , \dot{x} (s_1) \big) 
\dot{x}{}^{\nu} (s_1)  \dfrac{d\gamma ^{\mu}}{d \tau} ( \tau _1)
}{
g_{\rho \sigma} \big( x(s_2) ,\dot{x} (s_2) \big) 
\dot{x}{}^{\sigma} (s_2 ) \dfrac{d \tilde{\gamma}{}^{\rho}}{d {\tilde{ \tau}}} ( \tilde{\tau}{} _2)
}
\end{equation}
where $x^{\mu} (s)$ is a light ray connecting the emission event
at parameter value $s_1$ to the reception event at parameter value
$s_2$. For a derivation and a detailed discussion of this general
redshift formula in Finsler spacetimes we refer to a forthcoming article
by Hasse and Perlick \cite{HassePerlick2018}. The formula looks exactly
the same as the familiar redshift formula in a general-relativistic spacetime
(see, e.g., Straumann \cite{Straumann1984}), with the only modification
that now the metric depends  also on the tangent vector to the light ray.

The general redshift formula allows to test Finsler geometry on Earth, in
the Solar system and in cosmology. Details will be given in the above-mentioned
paper by Hasse and Perlick. Moreover, for 
cosmological redshift tests we refer to Hohmann and Pfeifer 
\cite{HohmannPfeifer2017}  who discuss the distance-redshift
relation in a cosmological Finsler spacetime based on the field
equation suggested by Pfeifer and Wohlfarth \cite{PfeiferWohlfarth2012}. 

%%%%%%%%%%%%%%%%%%%%%%%%%%%%%%%%%%%%%%%
\section{Conclusions}
Finsler geometry is a very natural generalisation of 
pseudo-Riemannian geometry and there are good
physical motivations for considering Finsler spacetime
theories. We have mentioned the Ehlers-Pirani-Schild
axiomatics and also the fact that a Finsler modification
of GR might serve as an effective theory of gravity that
captures some aspects of a (yet unknown) theory 
of Quantum Gravity. We have addressed the somewhat
embarrassing fact that there is not yet a general consensus 
on fundamental Finsler equations, in particular on 
Finslerian generalisations of the Dirac equation and of the 
Einstein equation, and not even on the question of which
precise mathematical definition of a Finsler spacetime 
is most appropriate in view of physics.
We have seen that the observational bounds on Finsler
deviations at the laboratory scale are quite tight. 
By contrast, at the moment
we do not have so strong limits on Finsler deviations
at astronomical or cosmological scales.

%%%%%%%%%%%%%%%%%%%%%%%%%%%%%%%%%%%%%%%
\section*{Acknowledgements}
We wish to thank Wolfgang Hasse, Manuel Hohmann and Christian Pfeifer 
for helpful discussions on Finsler geometry. Moreover, we acknowledge 
support from the DFG within the Research Training Group 1620 \emph{Models
of Gravity}.

%%%%%%%%%%%%%%%%%%%%%%%%%%%%%%%%%%%%%%%
%\section*{References}

%References are to be listed in the order cited in the text in Arabic
%numerals within square brackets. They can be
%referred to indirectly, e.g.~``$\ldots$
%in the statement \cite{beeson}.'' or used directly,
%e.g.~``$\ldots$ see [2] for examples.'' List references
%using the style shown in the following examples. For journal names,
%use the standard abbreviations.  Typeset references in 9 pt roman.

\end{document}